\documentclass[floatfix,reprint,superscriptaddress,amsmath,amssymb,aps,prl
]{revtex4-2}

\usepackage{graphicx}
\usepackage[dvipsnames]{xcolor}
\usepackage{pdfpages}
\usepackage{amssymb}
\usepackage{amsmath}
\usepackage{amsbsy}
\usepackage{color}
\usepackage{epstopdf}
\usepackage{bm}
\usepackage{xurl}
\usepackage{lipsum}
\usepackage{textcomp}

\usepackage{filecontents}
\makeatletter\@input{z2.tex}\makeatother

\graphicspath{{figures/}}

\usepackage{xr}
\makeatletter
\AtBeginDocument{\let\LS@rot\@undefined}
\newcommand*{\addFileDependency}[1]{
  \typeout{(#1)}
  \@addtofilelist{#1}
  \IfFileExists{#1}{}{\typeout{No file #1.}}
}
\makeatother
\newcommand*{\myexternaldocument}[1]{%
    \externaldocument[SM-]{#1}%
    \addFileDependency{#1.tex}%
    \addFileDependency{#1.aux}%
}
\myexternaldocument{si}

\begin{document} 
\title{Topological Signatures of Hardness and Structural Order in Network-Forming Materials}

\author{Yair Augusto Guti\'{e}rrez Fosado}
\thanks{Joint first author}
\thanks{corresponding author, yair.fosado@ed.ac.uk}
\affiliation{Centre for Science at Extreme Conditions and SUPA, School of Physics and Astronomy, University of Edinburgh, Peter Guthrie Tait Road, Edinburgh, EH9 3FD, UK}

\author{Ayobami Daramola}
\thanks{Joint first author}
\affiliation{Centre for Science at Extreme Conditions and SUPA, School of Physics and Astronomy, University of Edinburgh, Peter Guthrie Tait Road, Edinburgh, EH9 3FD, UK}

\author{Davide Marenduzzo}
\affiliation{Centre for Science at Extreme Conditions and SUPA, School of Physics and Astronomy, University of Edinburgh, Peter Guthrie Tait Road, Edinburgh, EH9 3FD, UK}

\author{Ciprian G. Pruteanu}
\affiliation{Centre for Science at Extreme Conditions and SUPA, School of Physics and Astronomy, University of Edinburgh, Peter Guthrie Tait Road, Edinburgh, EH9 3FD, UK}

\begin{abstract}
Understanding how the topology of network forming materials influences their physical properties remains a longstanding challenge. Here, we investigate the topology of experimentally compatible atomistic models of amorphous silica together with the crystalline polymorphs cristobalite and quartz. By comparing multiple atomistic models of amorphous silica that equally reproduce neutron-scattering data, we show that different microscopic descriptions can imply different topological interpretations of glass stability. To characterize the network beyond conventional geometric descriptors, we introduce the linking valence, which quantifies the average number of topological links per network loop. This topological descriptor separates silica into two distinct classes: the mechanically harder quartz exhibit values more than an order of magnitude larger than those of amorphous silica and cristobalite, despite their common tetrahedral building blocks. Spectral analysis of the loop-linking networks provides a complementary distinction, separating amorphous from crystalline phases and revealing differences in the long-range organization of topological constraints. These results establish topological linking as a new framework for connecting the structure and physical properties of network-forming materials.
\end{abstract}

\maketitle

\paragraph{Introduction --} Understanding how the topology of a network influences its physical properties remains a central challenge in the physics of network-forming materials. Silica provides the archetypal example. Amorphous silica~\cite{elliott1991medium,hannon2021basic,hobbs2008topological}, cristobalite, and quartz are all built from corner-sharing SiO$_4$ tetrahedra, yet exhibit markedly different densities, elastic responses, thermodynamic stabilities, and transformation pathways. Their structure has traditionally been characterized through pair distribution functions, coordination numbers, bond-angle distributions, and ring statistics, which successfully describe the local geometry and intermediate-range organization~\cite{micoulaut2003rings,sidebottom2019connecting,wondraczek2022advancing,liu2022vibrational,rouxel2007elastic}. By contrast, the topology of the network, in particular the mutual linking of rings, has received comparatively little attention, despite providing a natural description of nonlocal structural constraints that may influence rigidity and mechanical response.

A fundamental difficulty is that network topology cannot be directly inferred from diffraction experiments, which constrain pair correlations and therefore local geometry, but do not uniquely determine the underlying structure~\cite{alderman2022lead,wright1994neutron,keen1999local,bowron2008experimentally}. Consequently, distinct atomistic models can reproduce essentially the same structure factor while differing in their connectivity, ring organization, and topological constraints~\cite{hannon2021basic,liu2020searching,wu2012structure,liu2022vibrational,daramola2026densest,tucker2005refinement,zhou2021revealing,matsutani2024comparison}. This raises the question of whether network topology provides a robust structural descriptor that captures physically meaningful differences between phases beyond conventional geometric characterization.

To address this question, we analyze the topology of amorphous silica together with the crystalline polymorphs $\alpha$- and $\beta$-cristobalite and $\alpha$- and $\beta$-quartz. We find that the linking valence, defined as the average number of topological links per loop, separates the silica phases into two distinct topological classes. The mechanically harder quartz exhibit a linking valence more than an order of magnitude larger than that of amorphous silica and cristobalite, which instead belong to the same low-linking valence class. Spectral analysis of the loop-linking networks provides a complementary classification, separating the amorphous from the crystalline phases through differences in the long-range organization of their topological constraints. Together, these results show that distinct topological descriptors capture complementary aspects of network structure: linking valence provides a signature of mechanical rigidity, whereas spectral properties capture structural order.

\begin{figure*}[t]
	\centering
	\includegraphics[width=1.0\linewidth]{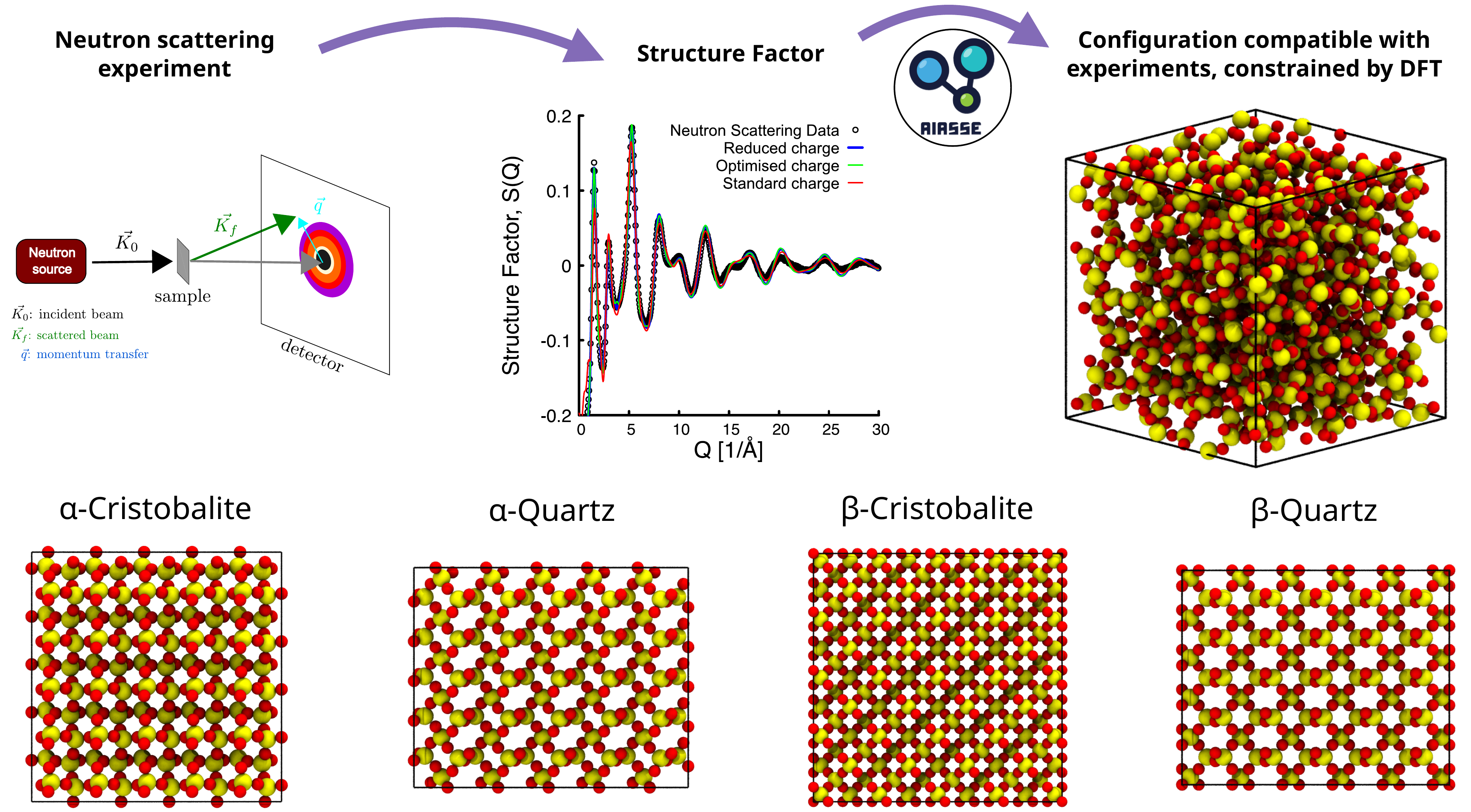}
\caption{\textbf{Experimentally constrained amorphous silica and crystalline reference structures}.
\textbf{Top:} AIASSE workflow used to generate amorphous silica configurations consistent with neutron-scattering data. EPSR refinement is performed using either standard ($q_{\mathrm{Si}}=+4$, $q_{\mathrm{O}}=-2$) or reduced ($q_{\mathrm{Si}}=+2$, $q_{\mathrm{O}}=-1$) fixed charges. For the optimized scheme, AIASSE iteratively updates the charges through DFT calculations, DDEC6 charge partitioning, and EPSR refinement until self-consistency is reached. A representative optimized-charge configuration is shown.
The central panel compares the experimental neutron-scattering structure factor with the amorphous silica models. All reproduce the main experimental features, with the optimized-charge model providing slightly improved agreement near the first sharp diffraction peak and at intermediate $Q$.
\textbf{Bottom:} Crystalline reference structures: $\alpha$- and $\beta$-cristobalite and $\alpha$- and $\beta$-quartz.
}
\label{fig:Silica-fit}
\vspace{-0.4cm}
\end{figure*}

\paragraph{Experimentally compatible configurations --} The amorphous silica configurations analyzed in this work were generated using the Ab initio Augmented Structure Solving Engine (AIASSE) framework~\cite{daramola2025aiasse}, which combines Empirical Potential Structure Refinement (EPSR)~\cite{soper2001tests} with self-consistent first-principles electronic-structure calculations~\cite{daramola2026densest}. The refinement is constrained by neutron scattering measurements, ensuring that all generated configurations reproduce the experimental structure factor, $F(Q)$, while permitting multiple atomistic realizations that are all consistent with the same experimental data used \cite{bowron2008experimentally} (see Fig.~\ref{fig:Silica-fit} and SI section~\ref{SM-sec:supp_methods} for details). Within this framework, we consider three electrostatic models, with all charges expressed in units of the elementary charge $e$: standard ($q_{\mathrm{Si}}=+4$, $q_{\mathrm{O}}=-2$), reduced ($q_{\mathrm{Si}}=+2$, $q_{\mathrm{O}}=-1$), and optimized ($q_{\mathrm{Si}}\approx+1.85$, $q_{\mathrm{O}}\approx-0.93$) obtained self-consistently using DDEC6 charge partitioning \cite{manz2016introducing}. Although all three charge schemes reproduce the experimental structure factor with comparable accuracy (Fig.~\ref{fig:Silica-fit}b), they produce subtle but systematic differences in the geometry and topology of the resulting networks (see SI, section \ref{SM-sec:conventionalStructure}). Each amorphous configuration contains $M=1500$ atoms (500 Si and 1000 O) under periodic boundary conditions at the experimental number density of $0.0664$ atoms\,\AA$^{-3}$.

The crystalline reference structures in Fig.~\ref{fig:Silica-fit} were generated from the corresponding Crystallographic Information Files. The unit cells were periodically replicated to generate simulation boxes containing approximately the same number of atoms as the amorphous configurations, thereby minimizing finite-size effects. The resulting systems contained $1725$ atoms for $\alpha$-cristobalite, $1500$ for $\alpha$-quartz, $2081$ for $\beta$-cristobalite, and $1525$ for $\beta$-quartz.

\begin{figure*}[t]
	\centering
	\includegraphics[width=1.0\linewidth]{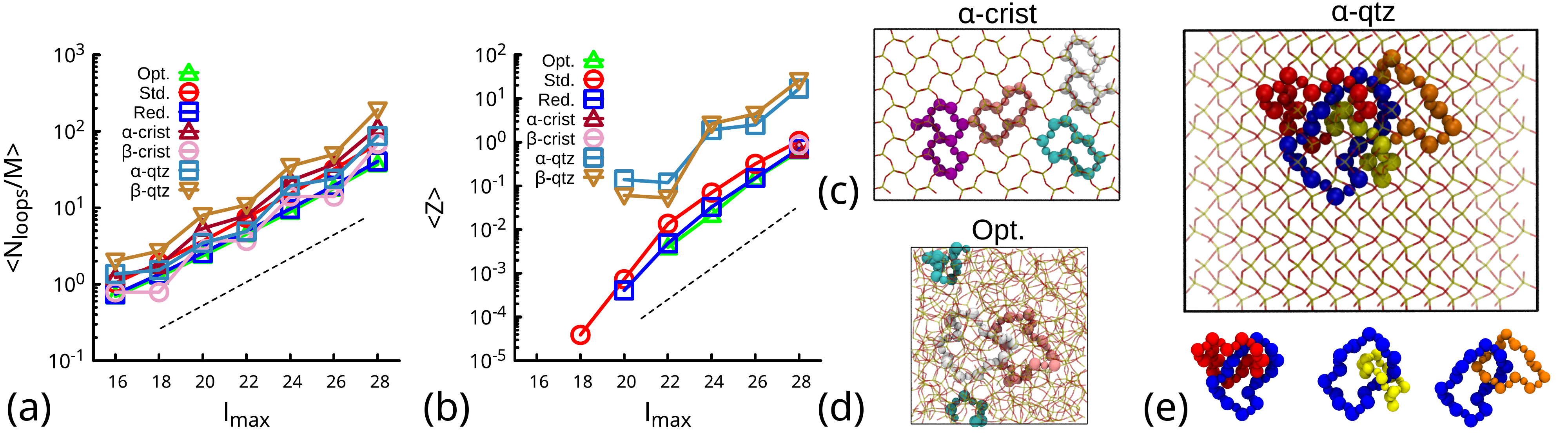}
\caption{\textbf{Topological analysis of silica networks.} \textbf{(a)} Average number of loops per particle, $N_{\rm loops}/M$, as a function of the maximum loop size $l_{\rm max}$ for amorphous silica, $\alpha$- and $\beta$-cristobalite, and $\alpha$- and $\beta$-quartz. The black dashed line shows the exponential scaling $N_{\rm loops}/M \propto \exp(0.35\,l_{\rm max})$. \textbf{(b)} Linking valence, $\mathcal{Z}=\mathcal{L}/N_{\rm loops}$, as a function of $l_{\rm max}$. The black dashed line shows the scaling $\mathcal{Z}\propto \exp(0.82\,l_{\rm max}$). \textbf{(c-e)} Representative examples of loops and mutually linked loops identified in different phases.
}
\label{fig:LocalTopology}
\vspace{-0.4cm}
\end{figure*}

\paragraph{Topological characterization of silica networks --} Motivated by recent studies on liquid–liquid phase separation in low- and high-density amorphous ices~\cite{Fosado2024,Sciortino2022}, as well as on DNA hydrogels exhibiting topologically nontrivial motifs such as Hopf links and trefoil knots~\cite{Fosado2025}, we investigate whether comparable topological complexity emerges in silica networks.

To characterize the topology of silica beyond its local tetrahedral geometry, each atomistic configuration is represented as a graph whose vertices correspond to atoms and whose edges represent Si-O bonds. Bonds are assigned using a cutoff distance $r_c$ defined by the first minimum of the silica radial distribution function. This construction naturally captures the connectivity of the corner-sharing tetrahedral network (see SI section~\ref{SM-sec:networkAnalysis}). 

The network topology is characterized through the closed loops of the graph. We identify all loops containing between $3$ and $l_{\rm max}$ vertices. Since exhaustive loop detection becomes computationally prohibitive for long loops, the maximum loop size was systematically increased from $l_{\rm max}=16$ to $28$, the largest computationally feasible value. As shown in Fig.~\ref{fig:LocalTopology}a, the number of loops per particle increases monotonically with $l_{\rm max}$ for all phases. Although the crystalline polymorphs generally contain more loops than the amorphous networks, loop abundance alone does not distinguish the mechanically different silica phases.

To quantify the mutual entanglement of loops, we compute the Gauss linking number between every pair of oriented loops $\gamma_i$ and $\gamma_j$,

\begin{equation}
Lk(\gamma_i, \gamma_j) = \frac{1}{4\pi} \oint_{\gamma_i} \oint_{\gamma_j} 
\frac{(\mathbf{r}_j - \mathbf{r}_i)}{\lvert \mathbf{r}_j - \mathbf{r}_i \rvert^3} \cdot 
\left( \text{d}\mathbf{r}_j \times \text{d}\mathbf{r}_i \right),
\label{eq:Lk}
\end{equation}

\noindent which counts the number of times two closed loops are mutually linked. Representative examples of loops and linked motifs identified in silica are shown in Figs.~\ref{fig:LocalTopology}c-e. The topology of the entire network is then characterized by the total linking, $\mathcal{L} = \sum_{i>j}^{N_{\text{loops}}} |Lk(\gamma_i, \gamma_j)|$, and by the linking valence $\mathcal{Z}=\mathcal{L}/N_{\text{loops}}$, which quantifies the average number of topological links per loop and, unlike conventional geometric descriptors, captures nonlocal topological constraints across the network. We therefore use the linking valence as the main topological descriptor throughout this work.

\begin{figure*}[t]
	\centering
	\includegraphics[width=1.0\linewidth]{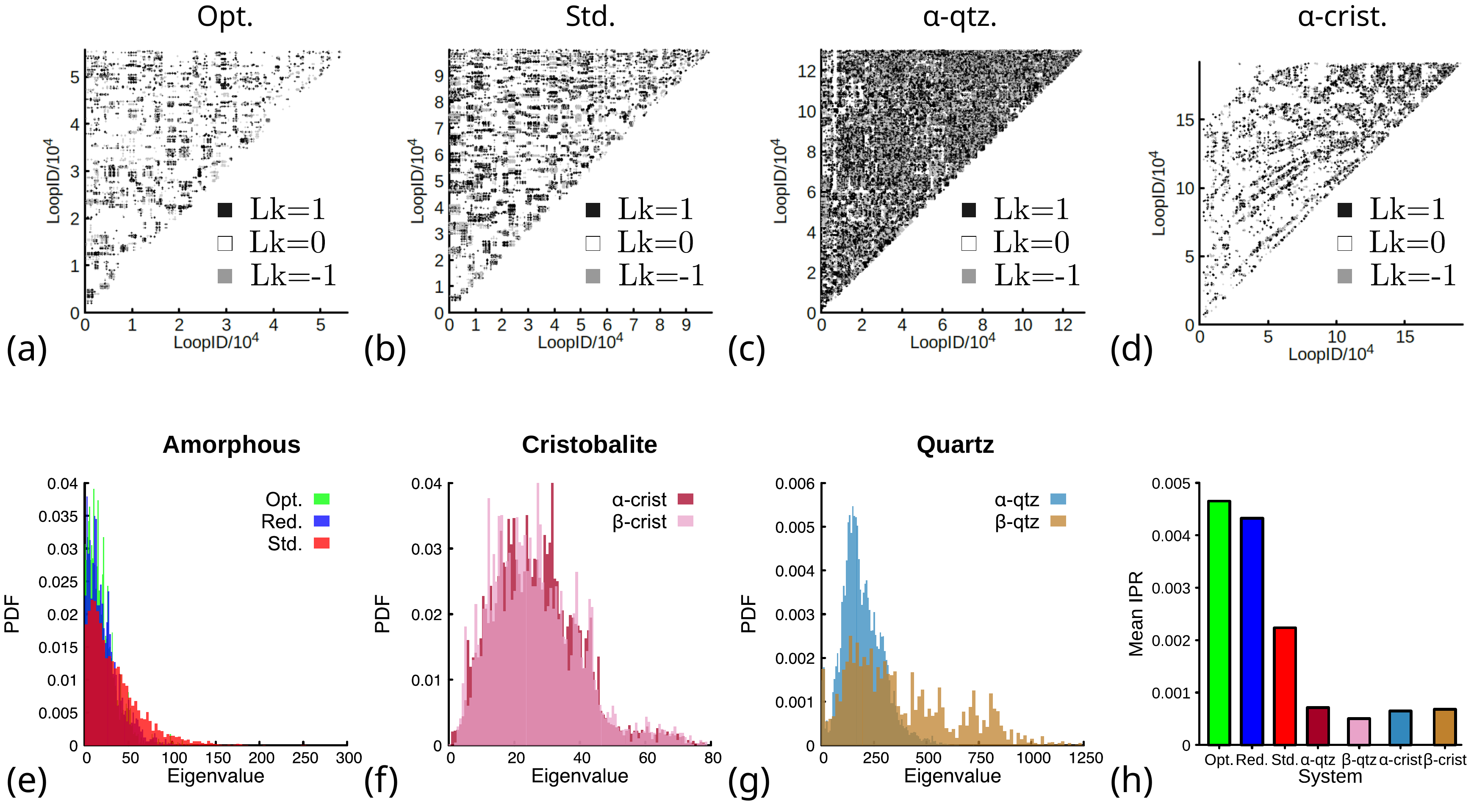}
\caption{\textbf{Global organization of topological constraints}. \textbf{(a)-(d)} Linking matrices $Lk_{ij}$ for different silica phases, all obtained using data with $l_{\rm max}=28$. \textbf{(e)} Laplacian eigenvalue spectra for the amorphous silica networks generated using optimized, reduced, and standard charge assignments. All three amorphous networks exhibit broad, monotonically decaying spectra. Interestingly, the decay rate of the amorphous network with standard-charge is significantly different, and larger, than those of the reduced and optimized charges, showing that quantum mechanics information leads to topologically different networks.
\textbf{(f)} Corresponding spectra for $\alpha$- and $\beta$-cristobalite, which display pronounced maxima at intermediate eigenvalues.
\textbf{(g)} Laplacian spectra for $\alpha$- and $\beta$-quartz, showing a similar peaked structure characteristic of the crystalline loop-linking networks.
\textbf{(h)} Mean inverse participation ratio (IPR) of the Laplacian eigenvectors for the amorphous and crystalline silica networks. The larger IPR values of the amorphous networks indicate stronger localization of graph eigenmodes, whereas the lower values for the crystalline polymorphs are consistent with more delocalized modes.
}    
\label{fig:GlobalTopology}
\vspace{-0.4cm}
\end{figure*}

\paragraph{Linking valence distinguishes mechanically distinct phases --} Figure~\ref{fig:LocalTopology}b shows the linking valence as a function of $l_{\rm max}$. For all values of $l_{\rm max}$ explored, $\alpha$- and $\beta$-quartz exhibit substantial linking, with $\langle\mathcal Z\rangle$ reaching $\approx16$-$25$, whereas amorphous silica and both cristobalite polymorphs remain weakly linked. The linking valence therefore separates the silica phases into two distinct topological classes: a low-linking class comprising amorphous silica and $\alpha$- and $\beta$-cristobalite, and a high-linking class formed by the quartz polymorphs. The observed sequence correlates with the experimentally-determined changes in Young's modulus and shear modulus (amorphous\cite{sorensen2024explaining,pabst2013elastic}, cristobalite\cite{pabst2013elastic,yeganeh1992elasticity, ji2018poisson}, quartz\cite{wang2015elasticity,peng2013mechanical}), and more importantly with the Mohs hardness of the material\cite{broz2006microhardness}, which all increase slightly from amorphous silica to cristobalite and then, more markedly, to quartz. Within the low-linking class, a further distinction emerges in the length scale at which topological linking appears: for both cristobalite polymorphs $\mathcal{Z}$ becomes nonzero only at $l_{\rm max}=28$, whereas amorphous silica exhibits finite linking over a broader range of loop sizes. Interestingly, within amorphous silica, this onset (at which $\mathcal{Z}\neq 0$) shifts systematically to larger loop sizes as the magnitude of the assigned charges is reduced, from $l_{\rm max}=18$ for the standard-charge model to $20$ and $22$ for the reduced- and optimized-charge models, respectively.

From a physical perspective, linked loops constitute nonlocal topological constraints that cannot be removed through continuous deformations of the network without breaking bonds. A high linking valence therefore reflects a strongly constrained network topology, whereas sparse linking is compatible with greater structural flexibility. This suggests that the marked differences in linking valence capture an aspect of network organization relevant to the mechanical response that is not apparent from conventional geometric characterization.
Interestingly, within the amorphous silica structures, the optimized-charge model has a linking valence comparable to that of $\alpha$-cristobalite, whereas the standard-charge model gives substantially stronger linking despite comparable agreement with diffraction data (Fig.~\ref{SM-fig:lkval28}). This shows that the inferred topological signature on glass stability depends on the underlying atomistic model (see SI section~\ref{SM-sec:networkAnalysis}).

We also observe a related trend in amorphous ice, where the high-density (HDAI) phase exhibits a larger linking valence than the low-density (LDAI) phase (Fig.~\ref{SM-fig:LkvalenceIce}), consistent with their distinct elastic response~\cite{Garku2022ice,Gromnitskaya2001}. This suggests that the connection between linking and mechanical response extends beyond silica. We emphasize, however, that linking valence is meaningful only for comparisons within the same material and should not be interpreted as an absolute measure of hardness across chemically distinct systems.

\paragraph{Linking matrices describe organization and localization of topological constraints --}
While the linking valence quantifies the average number of topological links per loop, it does not describe how those links are distributed throughout the network. To visualize this organization, we construct the linking matrix $Lk_{ij}$, whose elements correspond to the linking number between loops $i$ and $j$. Representative matrices for silica are shown in Fig.~\ref{fig:GlobalTopology}a-d. Amorphous silica exhibits sparse and heterogeneous linking matrices, reflecting that most loops are either unlinked or connected to only a few neighbors. By contrast, $\alpha$-quartz displays a much denser matrix, indicating that topological constraints are distributed throughout the network. Interestingly, $\alpha$-cristobalite remains weakly linked but exhibits a markedly more structured distribution of links than the amorphous phases.

To quantify this organization, we analyze the spectrum of the graph Laplacian, $L=D-A$, where $A$ is the adjacency matrix of the loop-linking graph and $D$ is the corresponding degree matrix. The graph Laplacian provides the canonical operator for characterizing global network connectivity, and its eigenvalue spectrum constitutes a compact fingerprint of the organization of loop-loop interactions (see SI section~\ref{SM-sec:spectralAnalysis} for details). As shown in Fig.~\ref{fig:GlobalTopology}e-g, amorphous silica displays monotonically decaying spectra for all charge schemes. In contrast, both cristobalite and quartz develop pronounced maxima at intermediate eigenvalues, revealing a characteristic connectivity scale absent from the amorphous networks. The inverse participation ratio (IPR) of the Laplacian eigenvectors (Fig.~\ref{fig:GlobalTopology}h) further reveals substantially stronger mode localization in amorphous silica than in the crystalline phases, indicating a more heterogeneous organization of topological constraints.
A comparison with amorphous ice reveals an analogous trend, with LDAI resembling amorphous silica and HDAI shifting toward the denser and more organized linking observed in $\alpha$-quartz (Fig.~\ref{SM-fig:spectralSiIce}).

Physically, the linking valence, Laplacian spectrum, and IPR probe the amount, organization, and localization of topological constraints, respectively. Together, they reveal distinct aspects of network topology: whereas linking valence distinguishes phases according to their mechanical character, the spectral properties of the linking network provide a complementary signature of crystalline versus amorphous organization. 

\paragraph{Discussion and conclusions --}
In summary, we have shown that topological linking provides a powerful way to classify networks arising in nature. First, the linking valence quantifies the abundance of nonlocal constraints and discriminates between mechanically hard and soft phases (Fig.~\ref{fig:LocalTopology}). Second, the Laplacian spectrum provides complementary information on how topological constraints are organized. In particular, the pronounced spectral maxima observed for crystalline silica and cristobalite 
compared with the broad, decaying spectra of amorphous silica and low-density amorphous ice, reveal distinct patterns of global connectivity. The corresponding inverse participation ratios further indicate stronger localization of topological modes in amorphous silica than in crystalline silica, while high-density amorphous ice exhibits an intermediate character (Figs.~\ref{fig:GlobalTopology} and \ref{SM-fig:spectralSiIce}). Together, these results show that the amount, organization, and localization of topological constraints provide complementary descriptors of network structure, capturing aspects related to both mechanical properties and glass-like structure. 

Our results further show that the topological picture depends on the atomistic description of the network. 
In amorphous silica, quantum-mechanically informed and standard charge models that are comparably compatible with diffraction data produce markedly different linking valences and mode localization (Figs.~\ref{fig:LocalTopology}b, \ref{SM-fig:lkval28} and \ref{fig:GlobalTopology}e). This highlights the importance of incorporating quantum-mechanical information when relating network topology to the physics and thermodynamics of amorphous materials. We also anticipate that loop-linking and spectral descriptors could prove useful for connecting nonlocal topology to physical properties in other complex networks, from filamentous biological systems to soft and active matter~\cite{charlton2025}.

\paragraph{Acknowledgements --} YAGF acknowledges support from the Physics of Life, UKRI/Wellcome (grant number EP/T022000/1–PoLNET3) and the contribution of the COST Action Eutopia, CA17139. For the purpose of open access, we have applied a Creative Commons Attribution (CCBY) licence to any author accepted manuscript version arising from this submission

\paragraph{Data availability --} Silica configurations and topology analysis codes are available at: \href{https://git.ecdf.ed.ac.uk/ygutier2/silica-topology}{\small{https://git.ecdf.ed.ac.uk/ygutier2/silica-topology}}

\bibliography{bibliography}

\end{document}


\title{Supplementary information: Topological Signatures of Hardness and Structural Order in Network-Forming Materials}

\author{Yair Augusto Guti\'{e}rrez Fosado}
\thanks{Joint first author}
\thanks{corresponding author, yair.fosado@ed.ac.uk}
\affiliation{Centre for Science at Extreme Conditions and SUPA, School of Physics and Astronomy, University of Edinburgh, Peter Guthrie Tait Road, Edinburgh, EH9 3FD, UK}

\author{Ayobami Daramola}
\thanks{Joint first author}
\affiliation{Centre for Science at Extreme Conditions and SUPA, School of Physics and Astronomy, University of Edinburgh, Peter Guthrie Tait Road, Edinburgh, EH9 3FD, UK}

\author{Davide Marenduzzo}
\affiliation{Centre for Science at Extreme Conditions and SUPA, School of Physics and Astronomy, University of Edinburgh, Peter Guthrie Tait Road, Edinburgh, EH9 3FD, UK}

\author{Ciprian G. Pruteanu}
\affiliation{Centre for Science at Extreme Conditions and SUPA, School of Physics and Astronomy, University of Edinburgh, Peter Guthrie Tait Road, Edinburgh, EH9 3FD, UK}

\maketitle

\section{Supplementary Methods}
\label{sec:supp_methods}

\subsection{Neutron-constrained structural refinement}

Structural models of amorphous silica were generated from neutron
total-scattering data using the \emph{Ab initio} Augmented Structure
Solving Engine (AIASSE)~\cite{daramola2025aiasse}. AIASSE couples
Empirical Potential Structure Refinement (EPSR)~\cite{soper2001tests}
with density-functional theory (DFT), allowing electronic-structure
information to be incorporated into diffraction-constrained structural
refinement.

Three electrostatic descriptions were considered: nominal ionic charges,
charges reduced by a factor of two, and DDEC6 charges obtained
self-consistently from DFT. All three models were refined independently
against the same experimental neutron structure factor using the same
SiO$_2$ composition, atomic number density, and initial reference
potential \cite{bowron2008experimentally}.

\subsection{Empirical Potential Structure Refinement}

The EPSR interaction was written as

\begin{equation}
U_{\alpha\beta}(r)
=
U_{\alpha\beta}^{\mathrm{ref}}(r)
+
U_{\alpha\beta}^{\mathrm{emp}}(r),
\label{eq:epsr_total}
\end{equation}

where the reference interaction consisted of Lennard--Jones and Coulomb
terms,

\begin{equation}
U_{\alpha\beta}^{\mathrm{ref}}(r)
=
4\varepsilon_{\alpha\beta}
\left[
\left(\frac{\sigma_{\alpha\beta}}{r}\right)^{12}
-
\left(\frac{\sigma_{\alpha\beta}}{r}\right)^{6}
\right]
+
\frac{q_{\alpha}q_{\beta}}
{4\pi\varepsilon_{0}r}.
\label{eq:epsr_ref}
\end{equation}

Unlike-species Lennard--Jones parameters were obtained using the
Lorentz--Berthelot combining rules,

\begin{equation}
\varepsilon_{\alpha\beta}
=
\sqrt{\varepsilon_{\alpha}\varepsilon_{\beta}},
\qquad
\sigma_{\alpha\beta}
=
\frac{\sigma_{\alpha}+\sigma_{\beta}}{2}.
\label{eq:LB}
\end{equation}

The initial reference-potential parameters were based on previous EPSR
modelling of silica~\cite{bowron2008experimentally}.

\begin{table}[htbp]
\centering
\caption{
Initial reference-potential parameters used for the EPSR refinement of
amorphous SiO$_2$. The charges correspond to the nominal ionic model.
}
\label{tab:params_si}
\begin{tabular}{lcccc}
\hline\hline
Atom &
Atomic fraction &
\(\varepsilon\) (kJ\,mol\(^{-1}\)) &
\(\sigma\) (\AA) &
\(q\) (\(e\)) \\
\hline
Si & 0.3333 & 0.80 & 0.76 & \(+4\) \\
O  & 0.6667 & 0.65 & 3.69 & \(-2\) \\
\hline\hline
\end{tabular}
\end{table}

Each system was first equilibrated using the reference potential. The
empirical contribution was then introduced and refined iteratively against
the experimental neutron structure factor. Sampling was continued until
the scattering function and principal structural observables were stable.

\subsection{AIASSE self-consistent charge refinement}

For the DFT-informed model, configurations sampled from the EPSR ensemble
were used as input for DFT calculations. Atomic charges were obtained from
the resulting electron density using the Density Derived Electrostatic and
Chemical method, sixth generation (DDEC6). Ensemble-averaged Si and O
charges were subsequently returned to the electrostatic term of the EPSR
reference potential.

The iterative procedure can be represented as

\begin{equation}
q^{(n)}
\;\xrightarrow{\mathrm{EPSR}}\;
\left\{\mathbf{R}\right\}^{(n)}
\;\xrightarrow{\mathrm{DFT/DDEC6}}\;
q^{(n+1)}.
\label{eq:aiasse_cycle}
\end{equation}

The EPSR--DFT cycle was repeated until the ensemble-averaged charges and
associated structural observables showed no systematic change between
successive iterations. The converged charges were then used in the final
AIASSE refinement.

\subsection{Electronic-structure calculations}

Electronic-structure calculations were performed with CP2K using the PBE
exchange--correlation functional and the Grimme D3 dispersion correction. Goedecker--Teter--Hutter pseudopotentials and DZVP-MOLOPT-SR-GTH Gaussian basis sets were employed, with an auxiliary plane-wave density cutoff of 400~$\rm{Ry}$. The valance density output was used in DDEC6 charge analysis attached to AIASSE.

The EPSR refinements employed a 1500-atom cell containing 500 Si and
1000 O atoms at an atomic number density of

\begin{equation}
\rho = 0.0664~\mathrm{atoms\,\AA^{-3}}.
\end{equation}

A smaller 300-atom cell containing 100 Si and 200 O atoms at the same
density was used for the DFT calculations.

\subsection{Charge models and statistical analysis}

The three electrostatic descriptions were

\begin{align}
\text{standard:}\qquad
&q_{\mathrm{Si}}=+4e,
& q_{\mathrm{O}}=-2e,
\\[3pt]
\text{reduced:}\qquad
&q_{\mathrm{Si}}=+2e,
& q_{\mathrm{O}}=-1e,
\\[3pt]
\text{AIASSE-optimised:}\qquad
&q_{\mathrm{Si}}=q_{\mathrm{Si}}^{\mathrm{DDEC6}},
& q_{\mathrm{O}}=q_{\mathrm{O}}^{\mathrm{DDEC6}}.
\end{align}

For the standard and reduced models, the charges were held fixed throughout
the refinement. For the AIASSE-optimised model, the charges were obtained
from the self-consistent EPSR--DFT procedure described above.

Twenty independent refinements were generated for each charge model.
Structural quantities were evaluated from the converged ensembles, and
reported uncertainties correspond to standard errors.


\section{Conventional structural descriptors of silica networks}
\label{sec:conventionalStructure}

Coordination numbers were obtained by integrating the corresponding partial
radial distribution functions to their first minima. Since the position of
the first minimum varied slightly between charge models, the cutoff was
determined independently for each refinement. Unless otherwise stated,
distributions were accumulated over 20\,000 configurations.

\subsection{First-shell coordination and two-body excess entropy}

\begin{figure}[t]
    \centering
    \includegraphics[width=0.9\linewidth]
    {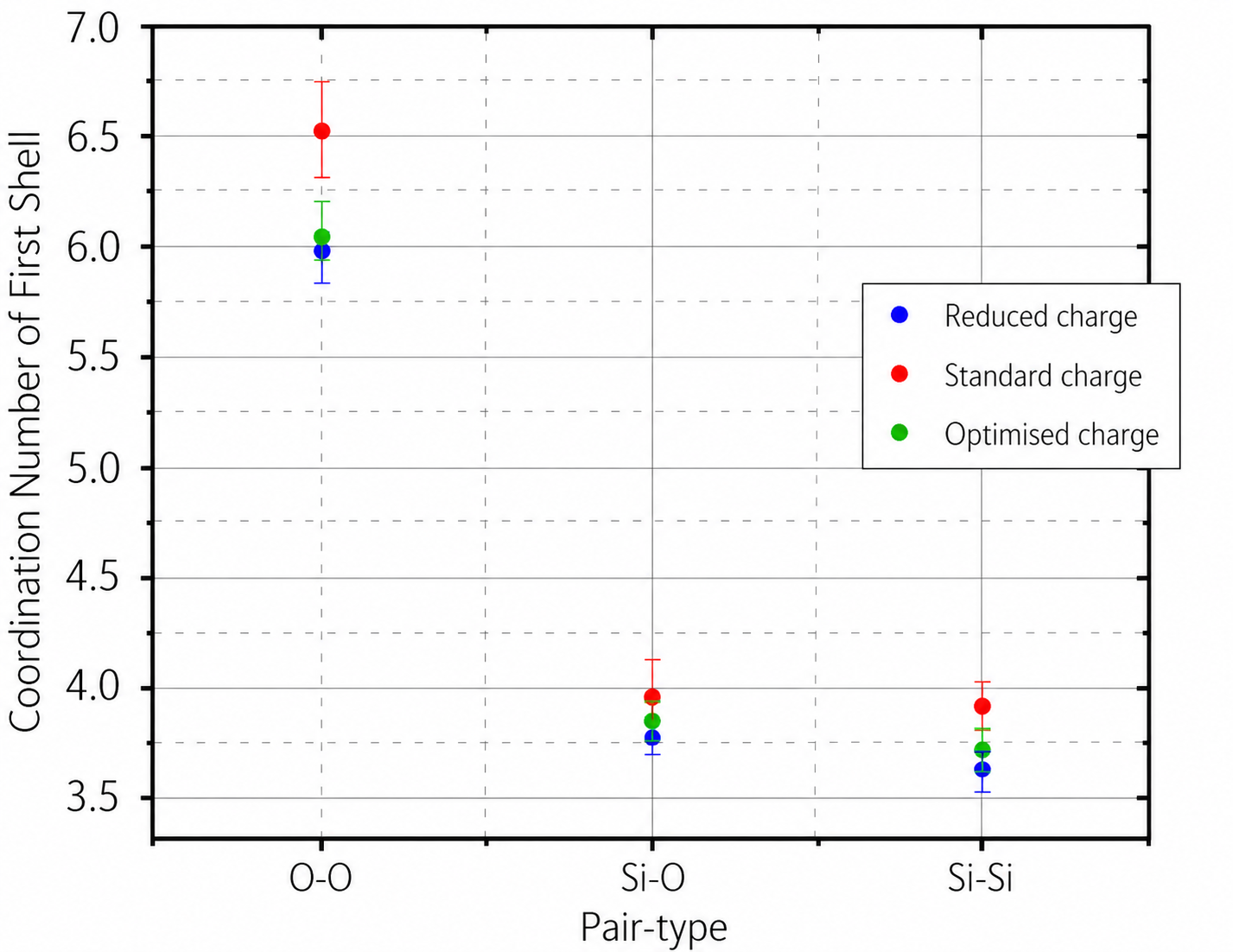}

    \vspace{0.4cm}

    \includegraphics[width=0.9\linewidth]
    {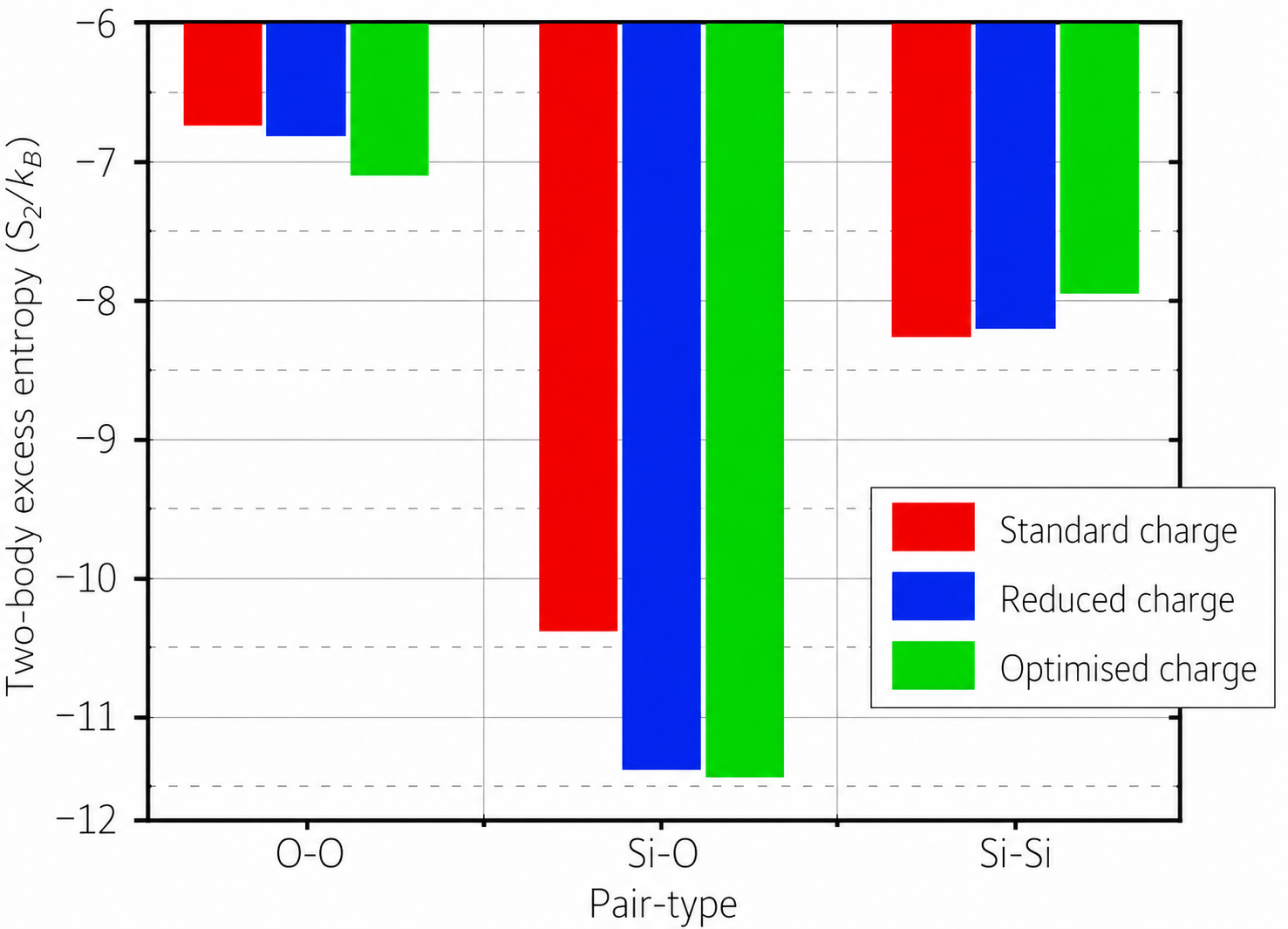}

    \caption{
    \textbf{First-shell coordination and pair-resolved two-body excess
    entropy of amorphous silica.}
    Top: first-shell coordination numbers for the O--O, Si--O, and Si--Si
    pairs. Bottom: corresponding pair contributions to the two-body excess
    entropy, \(S_{2}/k_{\mathrm{B}}\), for the standard, reduced, and
    AIASSE-optimised charge models.
    }
    \label{fig:coord_entropy}
\end{figure}

In Fig.~\ref{fig:coord_entropy} we show that the largest difference occurs in the O--O coordination, which decreases
from approximately 6.5 for the standard model to approximately 6.0 for
the reduced and AIASSE-optimised models. Si--O coordination remains close to four in all three refinements. The two-body excess entropy shows a corresponding pair-dependent response, with the Si--O contribution remaining largest in magnitude.

\begin{figure}[H]
    \centering
    \includegraphics[width=0.92\linewidth]{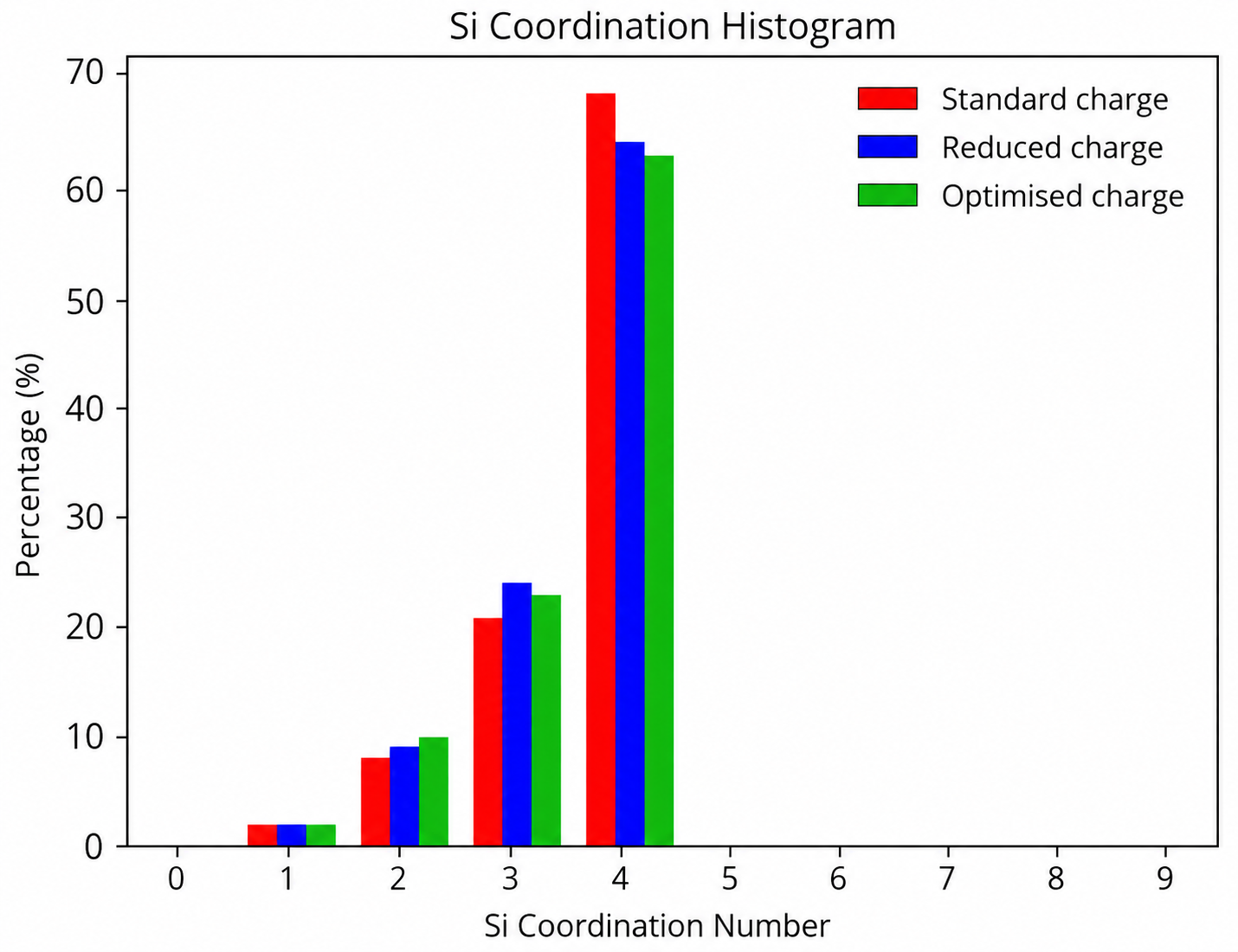}
    \caption{\textbf{Distribution of Si coordination numbers in amorphous silica.} Percentage of Si atoms with a given oxygen coordination number for the standard, reduced, and AIASSE-optimised charge models.
    }
    \label{fig:si_coordination}
\end{figure}

Fourfold-coordinated Si remains the dominant environment for all three
models, with comparatively small changes in the populations of the
lower-coordinated environments (see Fig.~\ref{fig:si_coordination}).

\subsection{Bond-angle distributions}

\begin{figure}[b]
    \centering
    \includegraphics[width=0.93\linewidth]
    {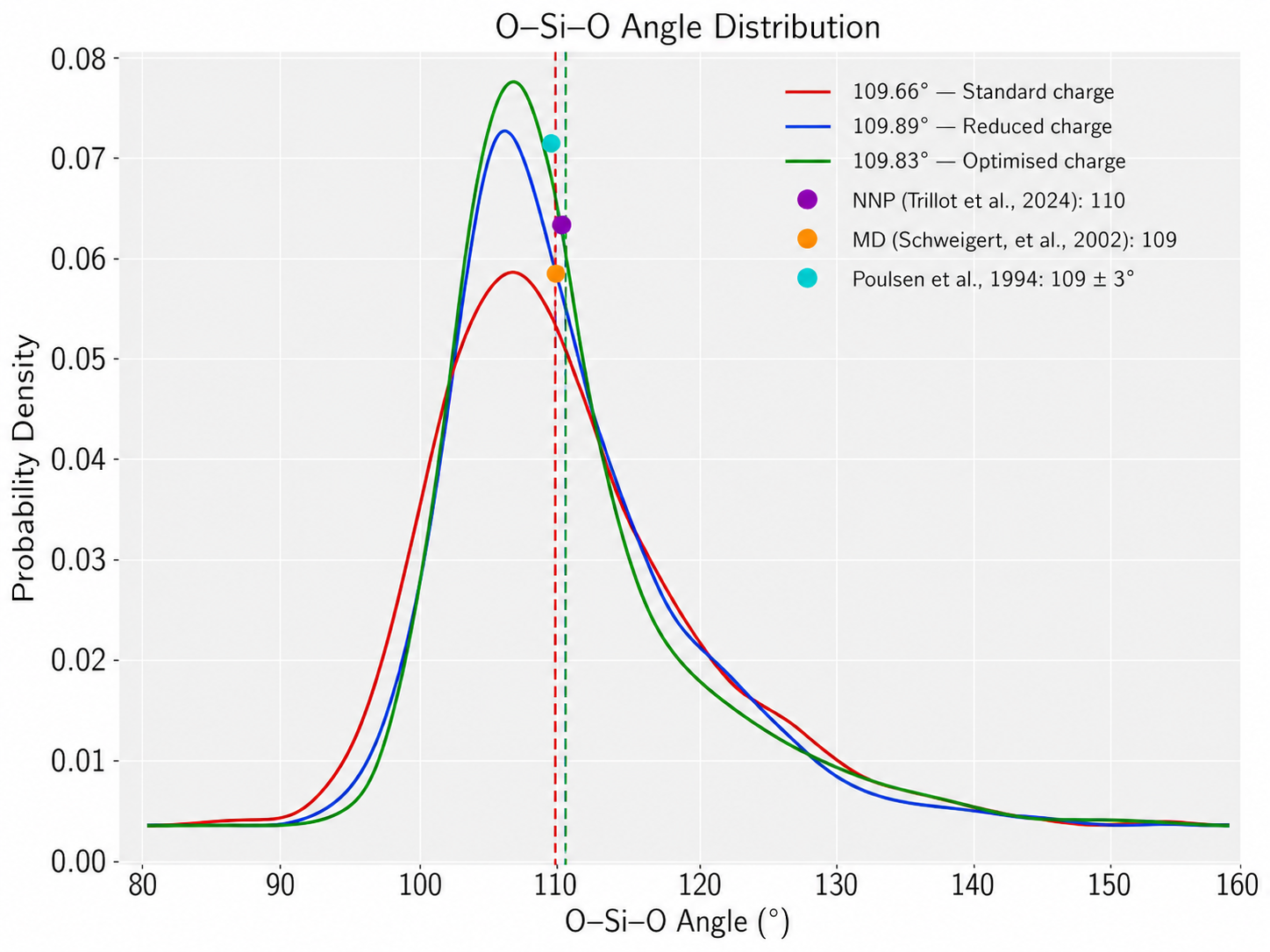}

    \vspace{0.5cm}

    \includegraphics[width=0.93\linewidth]
    {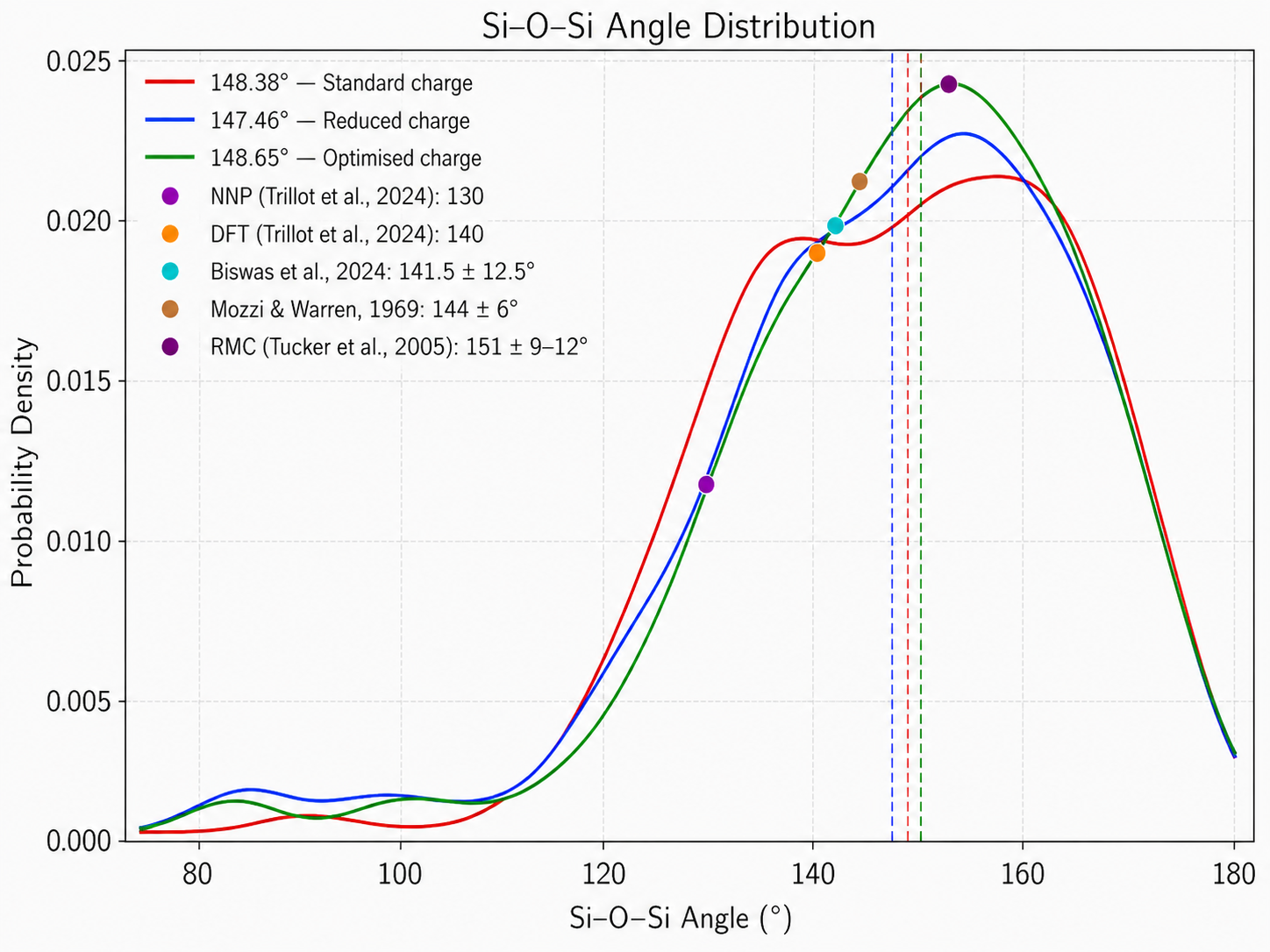}

    \caption{
    \textbf{Bond-angle distributions of amorphous silica.}
    Top: O--Si--O distributions for the standard, reduced, and
    AIASSE-optimised charge models, with mean values of
    \(109.66^{\circ}\), \(109.89^{\circ}\), and \(109.83^{\circ}\),
    respectively. Bottom: corresponding Si--O--Si distributions, with mean
    values of \(148.38^{\circ}\), \(147.46^{\circ}\), and
    \(148.65^{\circ}\). Literature values are included for comparison
    \cite{mozzi1969structure,tucker2005refinement,trillot2024elaboration}.
    }
    \label{fig:angle_distributions}
\end{figure}

In Fig.~\ref{fig:angle_distributions} we show that the mean O--Si--O and Si--O--Si angles vary only weakly with charge model. The main differences occur in the widths and shapes of the distributions, while all three refinements retain the characteristic tetrahedral geometry of amorphous silica.

\section{Electronic localisation in the AIASSE-refined structure}
\label{sec:electronic_localisation}

\begin{figure}[t]
    \centering
    \includegraphics[width=0.92\linewidth]{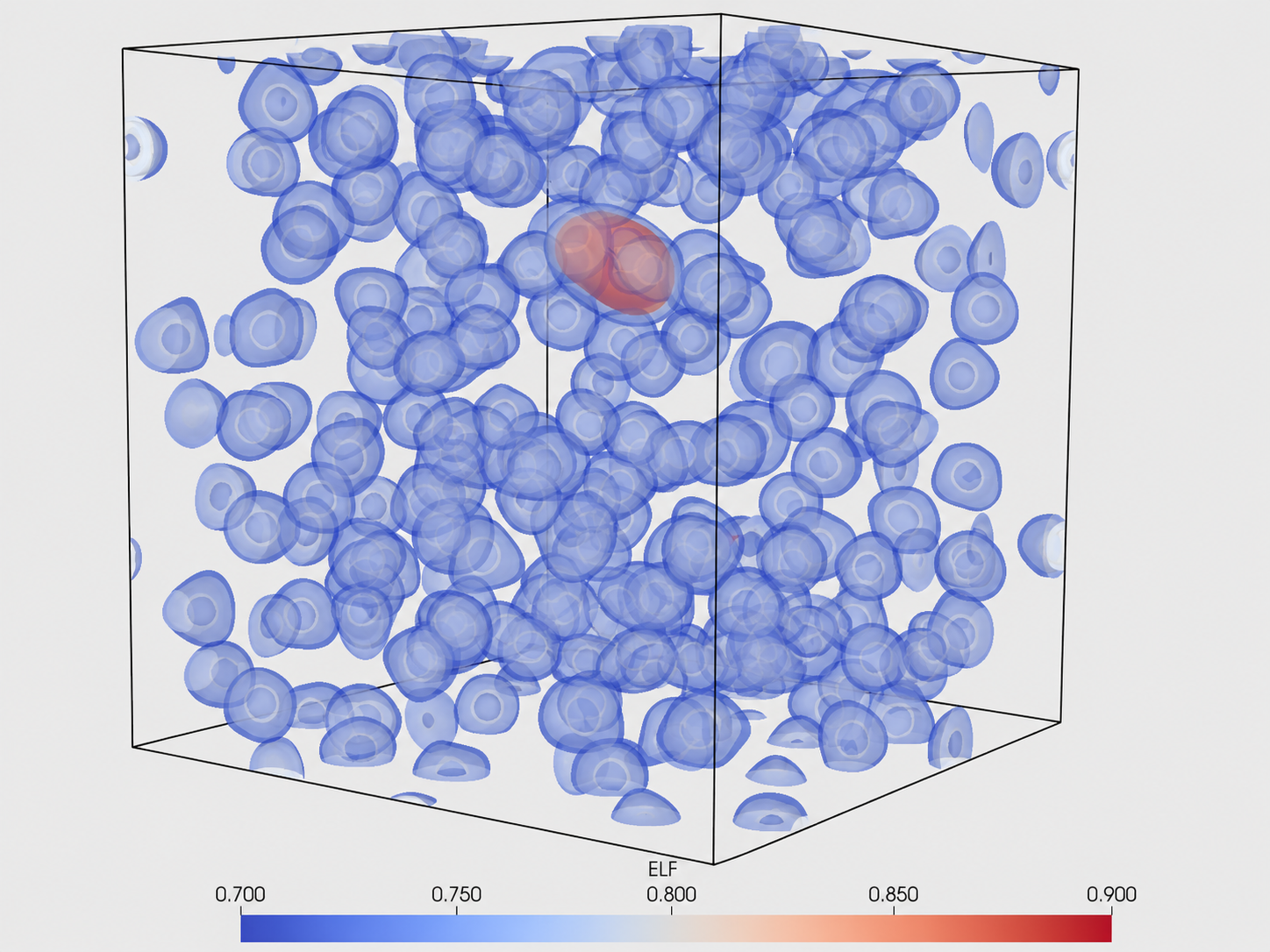}

    \caption{
    \textbf{Electron localisation function in amorphous silica.}
    Three-dimensional ELF isosurfaces for a representative configuration
    from the DFT-informed AIASSE ensemble. The colour scale spans ELF values
    from 0.70 to 0.90.
    }
    \label{fig:ELF}
\end{figure}

The ELF follows the local Si--O network and exhibits a strongly non-uniform
electronic-density distribution. Together with the DDEC6 analysis, this
shows substantial redistribution of electron density relative to a formal
ionic description. The DDEC6 values are therefore used here as effective
electrostatic charges derived from the DFT electron density rather than as
formal oxidation-state assignments.

\begin{figure*}[t]
	\begin{center}
		\includegraphics[width=1.0\textwidth]{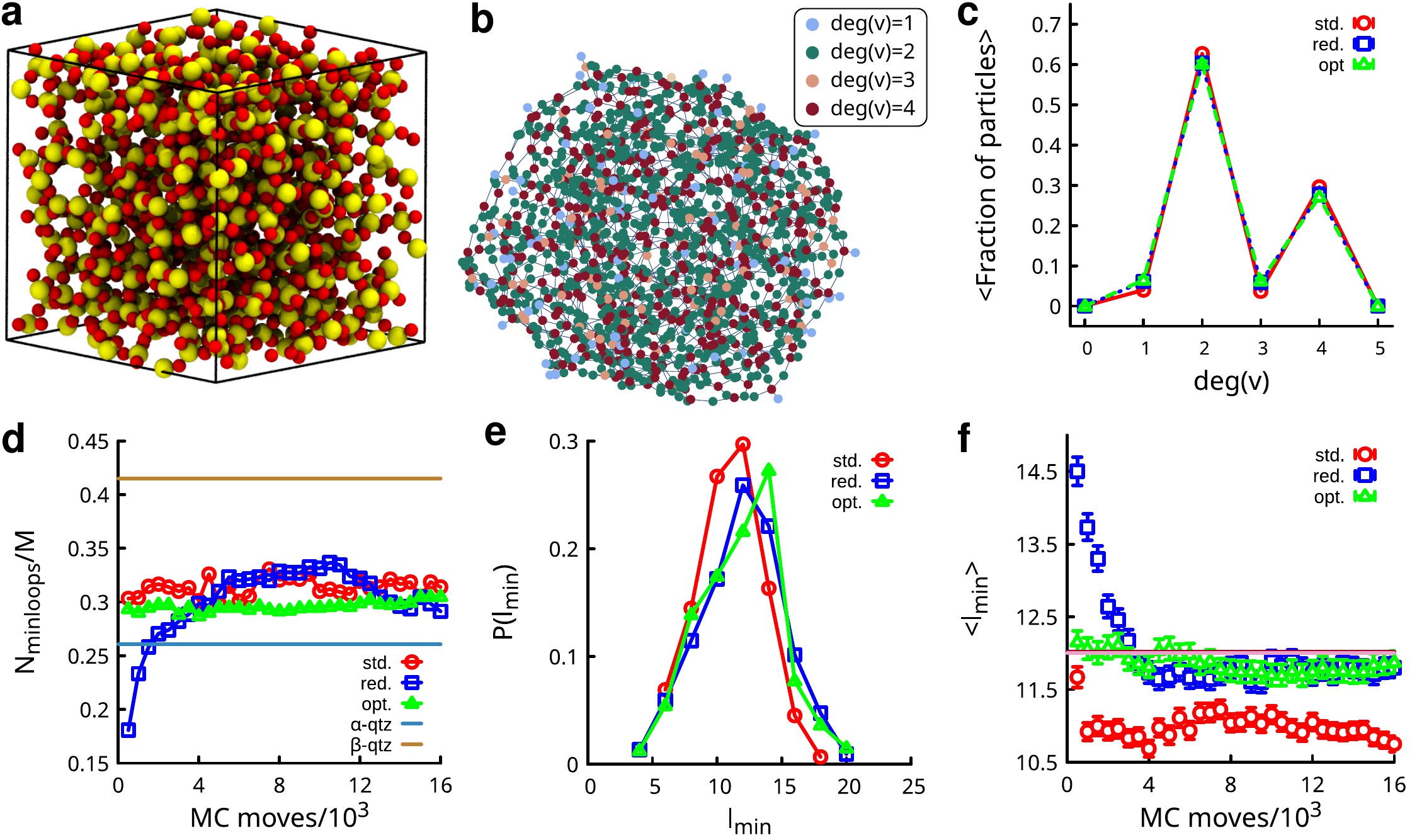}
        \caption{\textbf{Minimum-loop analysis in silica networks.}
        \textbf{a.} Representative silica configuration generated with AIASSE for the optimized-charge scheme.
        \textbf{b.} Network representation, where vertices denote atoms and edges represent bonds.
        \textbf{c.} Particle connectivity for the different charge schemes.
        \textbf{d.} Number of minimum loops per particle, $N_{\mathrm{minloops}}/M$, as a function of Monte Carlo moves for the different charge schemes. Straight lines indicate the corresponding values for the crystalline phases.
        \textbf{e.} Probability distribution of minimum-loop lengths, $P(l_{\mathrm{min}})$.
        \textbf{f.} Average minimum-loop length, $\langle l_{\mathrm{min}}\rangle$. Straight lines indicate the corresponding values for the crystalline phases. In all plots, lines connecting data points are guides to the eye.}
        \label{fig:minloops}
	\end{center}
\end{figure*}

\section{Amorphous ice configurations}
\label{sec:amorphousice}

The amorphous ice configurations analyzed in this work were obtained from the molecular dynamics simulations reported in Ref.~\cite{Fosado2024} and are publicly available in the associated repository~\href{https://git.ecdf.ed.ac.uk/ygutier2/amorphous-ice-topology}{\small{https://git.ecdf.ed.ac.uk/ygutier2/amorphous-ice-topology}}. Briefly, simulations were performed for $N=512$ rigid water molecules described by the TIP4P/2005 interaction potential~\cite{tip4p2005}, using the \textsc{GROMACS 2021.5} package~\cite{gromacs} in the isobaric ($NPT$) ensemble. Coulombic and Lennard-Jones interactions were calculated with a cutoff of $1.1$~nm, with long-range electrostatic interactions treated using the Particle-Mesh Ewald (PME) method. Temperature and pressure were controlled using the Nos\'e--Hoover thermostat~\cite{nose,hoover} and Berendsen barostat~\cite{berendsen1984molecular}, respectively, and the equations of motion were integrated with a time step of $1$~fs.

Amorphous ice was generated by cooling liquid water equilibrated at $T=300$~K to $T=100$~K at a rate of $1$~K/ns. The resulting configuration was then compressed at $T=100$~K at a rate of $0.01$~GPa/ns. In the present work, we analyze two configurations along this compression trajectory, at $P=0.01$~GPa and $P=2$~GPa, corresponding to the low-density (LDAI) and high-density (HDAI) amorphous states, respectively. Further details of the simulation protocol and the topology analysis for amorphous ices are provided in Ref.~\cite{Fosado2024}.

\section{Network and topology analysis}
\label{sec:networkAnalysis}

As described in the main text, to characterize the topology of silica beyond its local tetrahedral geometry, each atomistic configuration is represented as a graph whose vertices correspond to atoms and whose edges represent particles' connectivity. Bonds are assigned using a cutoff distance $r_c$ defined by the first minimum of the Si-O radial distribution function: any pair of particles separated by a distance $r\leq r_c$ is considered bonded and an edge is added between the corresponding vertices. This bonding criterion yields mainly Si-O contacts.

For amorphous silica, the same cutoff, $r_c=1.74$~\AA, was employed for all three charge schemes, whereas for the crystalline phases we used $r_c=1.65$, $1.60$, $1.69$, and $1.66$~\AA{} for $\alpha$-cristobalite, $\beta$-cristobalite, $\alpha$-quartz, and $\beta$-quartz, respectively. This construction naturally captures the connectivity of the corner-sharing tetrahedral network (see Fig.~\ref{fig:minloops}b,c). Statistical averages for the amorphous phases were obtained from 32 independent configurations sampled every 500 Monte Carlo steps, whereas each crystalline phase is represented by a single periodic structure.

For amorphous ice, the hydrogen-bond network is constructed by identifying pairs of water molecules whose oxygen atoms are separated by less than $2.2$~\AA{} and for which the O--H--O angle is smaller than $30^\circ$~\cite{luzar1996hydrogen}. Each water molecule is then represented as a vertex and each hydrogen bond as an edge.

To characterize the network structure beyond local connectivity, we analyze loops in the graph representation (Fig.~\ref{fig:minloops}b), defined as closed paths with no repeated vertices or edges. We consider two classes of loops. First, we identify the minimum loops, corresponding to the shortest loops found locally in the network; these loops exhibit no topological linking in any of the silica or ice phases considered. Second, we identify all loops containing up to a maximum number of particles, $l_{\rm max}$. This latter analysis forms the basis of the results presented in the main text and reveals substantial topological linking between loops.

\subsection{Minimum loops}
Here, we assess the influence of charge assignment on the resulting network structure. Starting from silica configurations generated with AIASSE under the constraint of reproducing the experimental structure factor (see Fig.~\ref{M-fig:Silica-fit} and Fig.~\ref{fig:minloops}a), we compare networks obtained using fixed standard ($q_{\mathrm{Si}}=+4$, $q_{\mathrm{O}}=-2$) and reduced ($q_{\mathrm{Si}}=+2$, $q_{\mathrm{O}}=-1$) charges with those obtained using optimized charges ($q_{\mathrm{Si}}=+1.853$, $q_{\mathrm{O}}=-0.9265$). We find subtle but systematic differences in the resulting network geometry.

In contrast to the analysis presented in the main text, here we focus on minimum loops. For each vertex, we identify all loops passing through it and retain those with the smallest number of vertices, denoted by $l_{\mathrm{min}}$. Figure~\ref{fig:minloops}d shows the total number of minimum loops, normalized by the number of particles, $N_{\mathrm{minloops}}/M$, as a function of Monte Carlo moves. We further compute the distribution $P(l_{\mathrm{min}})$ (Fig.~\ref{fig:minloops}e) and its first moment $\langle l_{\mathrm{min}}\rangle$ (Fig.~\ref{fig:minloops}f). On average, minimum loops in the standard-charge case are shorter ($\langle l_{\mathrm{min}}\rangle=10.9\pm2.5$) than in the reduced-charge ($\langle l_{\mathrm{min}}\rangle=11.8\pm3.1$) and optimized-charge ($\langle l_{\mathrm{min}}\rangle=11.8\pm3.3$) cases. A complementary analysis of the radius of gyration shows that minimum loops in the standard-charge scheme are not only shorter but also more spatially compact.

For comparison, we perform the same analysis for the crystalline silica phases. $\alpha$-quartz exhibits the lowest density of minimum loops, with $N_{\mathrm{minloops}}/M=391/1500=0.26$. The remaining crystalline phases exhibit higher loop densities than the amorphous networks, with $N_{\mathrm{minloops}}/M=688/2081=0.33$ for $\beta$-cristobalite, $633/1525=0.41$ for $\beta$-quartz, and $932/1725=0.54$ for $\alpha$-cristobalite. In all crystalline phases, the average minimum loop length is close to $l_{\mathrm{min}}=12$, lying above the average value for the standard-charge amorphous silica and close to those of the reduced- and optimized-charge schemes.

These results show that, while the local connectivity remains largely unchanged (Fig.~\ref{fig:minloops}c), the charge assignment influences the medium-range organization of the amorphous network (Figs.~\ref{fig:minloops}d-f). In particular, relative to the standard-charge case, the distributions for the reduced- and optimized-charge schemes are shifted toward larger minimum-loop sizes, consistent with their larger values of $\langle l_{\mathrm{min}}\rangle$.

\subsection{All loops up to a maximum length}
We now describe the loop analysis used for the results presented in the main text. In contrast to the minimum-loop analysis discussed above, here we consider all loops up to a prescribed maximum length, $l_{\rm max}$. For each vertex $i$, with $i\in[1,M]$ and $M$ the total number of vertices in the graph, we identify all loops containing between 3 and $l_{\rm max}$ particles. Since exhaustive loop detection becomes computationally prohibitive for long loops, $l_{\rm max}$ was systematically increased from 16 to 28 for the silica networks, with 28 being the largest value computationally feasible for all phases considered.

For a fixed $l_{\rm max}$, the search proceeds sequentially through the vertices of the graph. After identifying all loops containing vertex $i$, that vertex is removed from the graph before continuing the search from vertex $i+1$. This procedure avoids counting the same loop multiple times and substantially reduces the computational cost. To prevent paths that close only through periodic boundary conditions from being incorrectly identified as loops, the coordinates of each candidate loop are reconstructed using the minimum-image convention (MIC). We additionally require the end-to-end distance of the reconstructed path to satisfy $R<r_c$, ensuring that its terminal vertices are connected by a bond. Repeating this procedure over all vertices yields the complete set of distinct loops containing at most $l_{\rm max}$ particles, with the total number of loops given by
\begin{equation}
N_{\rm loops}(l_{\rm max})=\sum_{i=1}^{M}N_i(l_{\rm max}),
\end{equation}
where $N_i(l_{\rm max})$ is the number of loops identified when processing vertex $i$.

For amorphous ice, we follow an analogous procedure, identifying all loops containing between 3 and 13 water molecules, with 13 being the largest computationally feasible value. These correspond to loops containing between 6 and 26 particles.

In Figs.~\ref{M-fig:LocalTopology}a,f of the main text, we report the total number of loops obtained using this procedure, normalized by the total number of vertices in the graph, as a function of $l_{\rm max}$ for silica and ices, respectively.

\begin{figure}[t]
    \centering
    \includegraphics[width=0.92\linewidth]{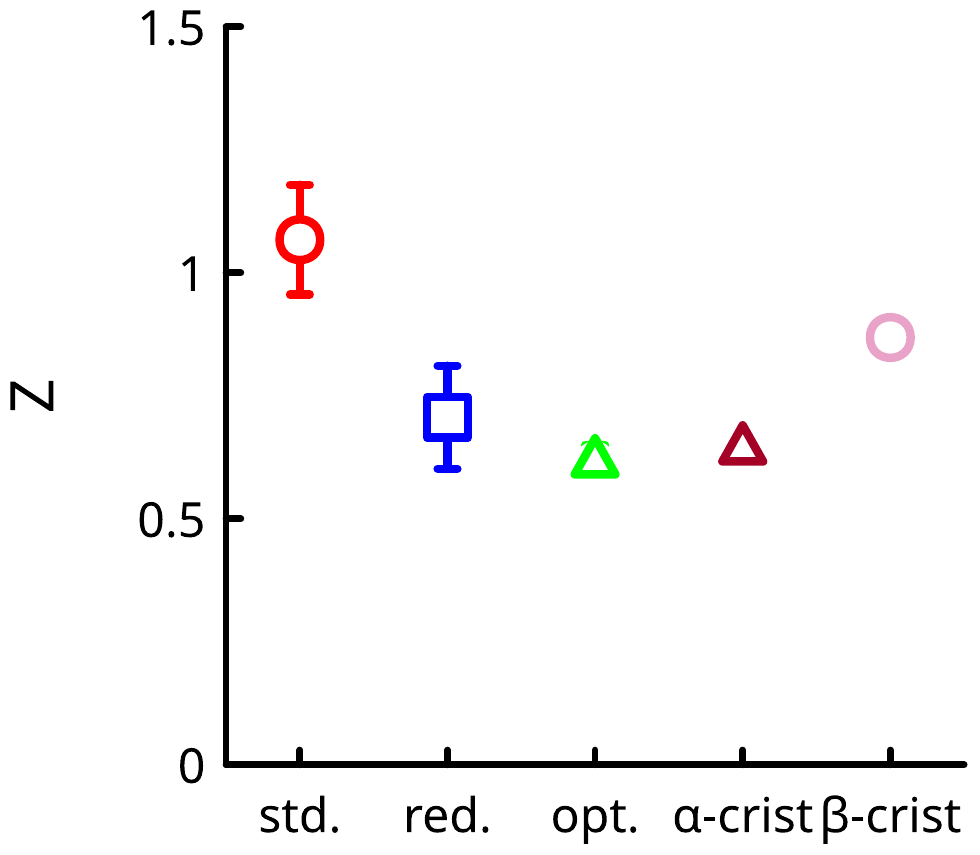}
    \caption{\textbf{Linking valence at $l_{\rm max}=28$ in silica networks.}
    Linking valence $\mathcal{Z}$ for the amorphous silica and crystalline silica phases at fixed $l_{\rm max}=28$.}
    \label{fig:lkval28}
\end{figure}

\subsection{Topological linking}
\begin{figure*}[t]
	\begin{center}
		\includegraphics[width=1.0\textwidth]{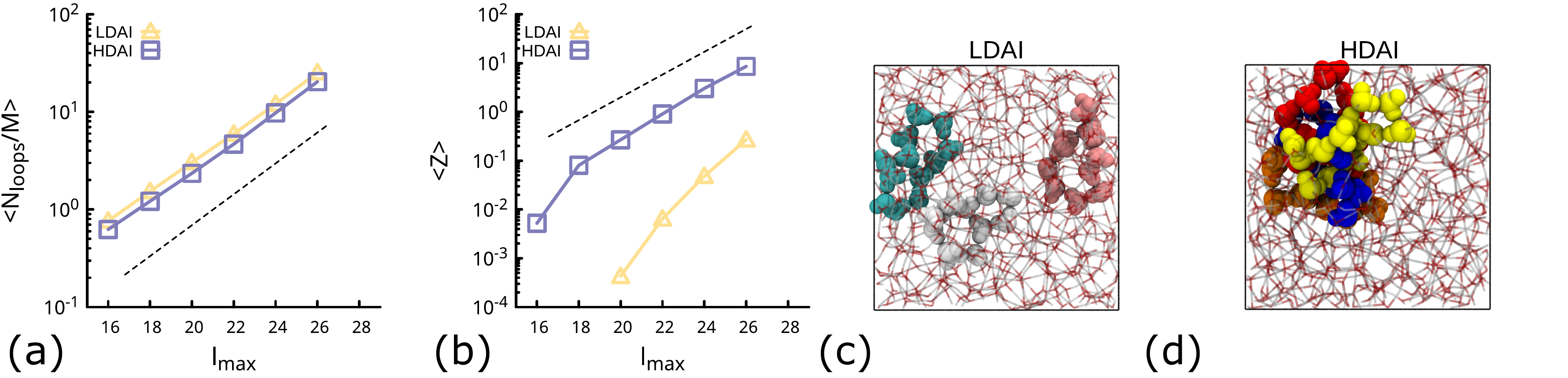}
\caption{\textbf{Topological analysis of amorphous ice networks.} \textbf{(a)} Average number of loops per particle, $N_{\rm loops}/M$, as a function of the maximum loop size $l_{\rm max}$ for for low-density (LDAI) and high-density (HDAI) amorphous ice. The black dashed line shows the exponential scaling $N_{\rm loops}/M \propto \exp(0.36\,l_{\rm max})$. \textbf{(b)} Linking valence, $\mathcal{Z}=\mathcal{L}/N_{\rm loops}$, as a function of $l_{\rm max}$. The black dashed line shows the scaling $\mathcal{Z}\propto \exp(0.537\,l_{\rm max}$). \textbf{(c-d)} Representative examples of loops and mutually linked loops identified in the different phases
}
        \label{fig:LkvalenceIce}
	\end{center}
\end{figure*}

To characterize non-trivial topological linking between loops in the network, we compute the linking number $Lk(\gamma_i,\gamma_j)$ between pairs of closed, oriented curves $\gamma_i$ and $\gamma_j$ using the Gauss linking integral,
\begin{equation}
Lk(\gamma_i,\gamma_j)
=
\frac{1}{4\pi}
\oint_{\gamma_i}
\oint_{\gamma_j}
\frac{\mathbf{r}_j-\mathbf{r}_i}
{\lvert\mathbf{r}_j-\mathbf{r}_i\rvert^3}
\cdot
\left(
\mathrm{d}\mathbf{r}_j\times\mathrm{d}\mathbf{r}_i
\right),
\end{equation}
where $\mathbf{r}_i$ and $\mathbf{r}_j$ denote the positions along the curves $\gamma_i$ and $\gamma_j$, respectively. Each curve is constructed from the sequence of vertices and edges defining the corresponding loop in the network.

Because the underlying graphs are undirected, there is no preferred orientation for the loops, and their orientations are therefore assigned arbitrarily. Consequently, the sign of $Lk$ carries no physical information in the present analysis, and averaging the signed linking number over all loop pairs yields a value close to zero. To quantify the overall degree of topological linking independently of orientation, we therefore define the total absolute linking number as
\begin{equation}
\mathcal{L}
=
\sum_{i>j}^{N_{\rm loops}}
\left|Lk(\gamma_i,\gamma_j)\right|.
\label{eq:totalLk}
\end{equation}
Thus, $\mathcal{L}$ measures the total amount of pairwise linking among all loops in the network. We further define the linking valence as $\mathcal{Z}=\mathcal{L}/N_{\rm loops}$, which measures the amount of topological linking normalized by the number of loops.

In Figs.~\ref{M-fig:LocalTopology}b of the main text, we report the linking valence as a function of $l_{\rm max}$ for all silica networks. For completeness, we also compare the linking valence across the different silica phases at fixed $l_{\rm max}=28$ in Fig.~\ref{fig:lkval28}. For amorphous silica, we find $\mathcal{Z}=1.07$, $0.71$, and $0.62$ for the standard, reduced, and optimized charge schemes, respectively. For $\alpha$- and $\beta$-cristobalite, $\mathcal{Z}=0.65$ and $0.87$, respectively, whereas for $\alpha$- and $\beta$-quartz, $\mathcal{Z}=16.8$ and $26.1$. These values highlight the pronounced separation between the weakly linked amorphous and cristobalite phases and the strongly linked quartz polymorphs.

\subsection{Topological interpretation of glass stability}
Whether the long-term stability of network glasses originates primarily from configurational entropy or from topological constraints remains a longstanding open question~\cite{mauro2014statistical,micoulaut2002network,han2020atomistic,toledo2019competition,sidebottom2025network,feng2022entropy,zaccarelli2007}. Our results in the previous section show that the answer depends sensitively on the atomistic description of the amorphous network, even when different models reproduce the same experimental structure factor. Using the quantum-mechanically informed charge assignment, optimized amorphous silica exhibits a linking valence ($\mathcal{Z}=0.62$, at $l_{\rm max}=28$) that is nearly identical to that of the stable $\alpha$-cristobalite phase ($\mathcal{Z}=0.65$). This close agreement indicates that both phases possess essentially the same degree of nonlocal topological constraint, suggesting that their markedly different thermodynamic stability cannot be attributed to topological linking alone and is therefore more likely to arise from configurational entropy or other energetic contributions. By contrast, the standard-charge model predicts a linking valence almost twice as large ($\mathcal{Z}\approx1.1$), implying a substantially more constrained amorphous network than the crystalline phase. Within such a description, the enhanced stability of the glass could instead be interpreted as arising from stronger topological constraints. These observations demonstrate that experimentally compatible atomistic models can lead to qualitatively different physical interpretations of the origin of glass stability, highlighting the importance of incorporating quantum-mechanical information when relating network topology to the thermodynamics of amorphous silica.

\subsection{Linking valence in amorphous ice}
\label{SM-sec:LkAmorphousice}
Here, we perform for low-density (LDAI) and high-density (HDAI) amorphous ice the same topological analysis presented for silica in Fig.~\ref{M-fig:LocalTopology} of the main text. Figure~\ref{fig:LkvalenceIce} shows the number of loops per atom and the corresponding linking valence as a function of $l_{\rm max}$. While the number of loops is comparable between the two amorphous phases, their linking valences differ markedly: HDAI exhibits substantially stronger linking than LDAI over the range of $l_{\rm max}$ considered. This enhanced linking in HDAI is consistent with the distinct elastic response of the two amorphous phases~\cite{Garku2022ice,Gromnitskaya2001} and provides an analogous trend to that observed between the mechanically distinct silica phases.

\begin{figure*}[t]
	\begin{center}
		\includegraphics[width=1.0\textwidth]{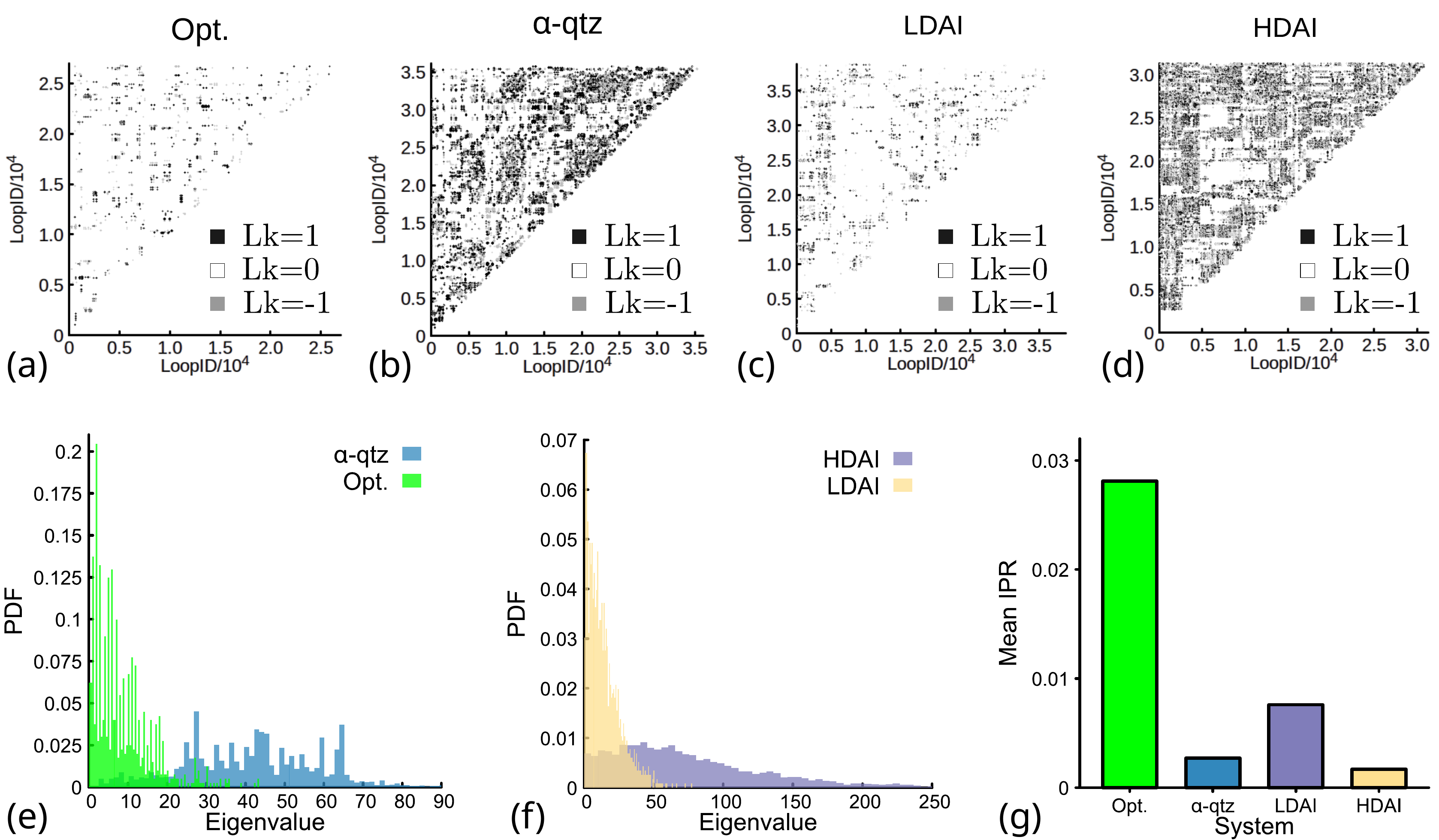}
        \caption{\textbf{Global organization of topological constraints.}
Linking matrices $Lk_{ij}$ for (a) amorphous silica (optimized charges), (b) $\alpha$-quartz, (c) low-density amorphous ice (LDAI), and (d) high-density amorphous ice (HDAI), all obtained using data with $l_{\rm max}=26$. Matrix element $(i,j)$ gives the Gauss linking number between loops $i$ and $j$, while zero entries correspond to unlinked loop pairs. (e,f) Probability density of the graph Laplacian eigenvalues for (e) silica and (f) amorphous ice. (g) Inverse participation ratio (IPR) of the Laplacian eigenvectors for the different networks analyzed.
}
        \label{fig:spectralSiIce}
	\end{center}
\end{figure*}

\section{Spectral analysis of loop-linking networks}
\label{sec:spectralAnalysis}

The loop-linking matrices naturally define undirected graphs in which each node corresponds to a primitive loop and an edge is drawn between two nodes whenever the corresponding loops possess a non-zero Gauss linking number. Spectral graph theory provides a convenient framework for characterizing the global organization of such networks through the eigenvalues and eigenvectors of graph operators \cite{Chung1997,Newman2018}.

For each linking graph one can construct the (unsigned) adjacency matrix $A$, with elements
\begin{equation}
A_{ij}=
\begin{cases}
1,& |Lk_{ij}|>0,\\
0,& \mathrm{otherwise},
\end{cases}
\end{equation}
where $Lk_{ij}$ denotes the Gauss linking number between loops $i$ and $j$. The corresponding degree matrix is diagonal,
\begin{equation}
D_{ii}=\sum_j A_{ij},
\end{equation}
and the combinatorial graph Laplacian is defined as
\begin{equation}
L=D-A.
\end{equation}
The Laplacian provides a compact description of the global connectivity of a graph and plays a central role in graph partitioning, diffusion processes and network robustness~\cite{Chung1997,Newman2018}.

\subsection{Laplacian spectra}

The eigenvalues of the graph Laplacian,
\begin{equation}
L\mathbf{v}_\alpha=\lambda_\alpha\mathbf{v}_\alpha,
\end{equation}
constitute a spectral fingerprint of the loop-linking network. The smallest eigenvalue is identically zero, while the remaining eigenvalues encode the connectivity and organization of the graph. In particular, sparse and weakly connected networks typically display broad spectra with many low-lying eigenvalues, whereas more homogeneous and strongly connected graphs develop characteristic spectral bands or maxima \cite{Chung1997,Spielman2007}. 

\subsection{Inverse participation ratio}

To quantify the spatial localization of the eigenmodes of the Laplacian matrix, we compute the inverse participation ratio (IPR) of each normalized eigenvector $\mathbf{v}_\alpha$,
\begin{equation}
\mathrm{IPR}_\alpha=
\sum_{i=1}^{N}
|v_{\alpha,i}|^4.
\end{equation}
The IPR measures the fraction of nodes contributing significantly to a given eigenmode. Extended modes distributed over the entire graph satisfy $\mathrm{IPR}\sim N^{-1}$, whereas localized modes concentrated on a small subset of nodes exhibit substantially larger values \cite{Bell1970,Evers2008}. In the present context, large IPR values indicate that the corresponding topological mode is localized on a limited number of strongly interacting loops, while small IPR values reflect collective organization over a large fraction of the loop-linking network.

\section{Spectral signatures of silica and amorphous ice loop-linking networks at $l_{\mathrm{max}}=26$}
\label{sec:topologySiIce}

In this Section we discuss the main topological features of silica (amorphous-optimized and $\alpha$-quartz) and amorphous-ice (HDAI and LDAI) networks. Representative linking matrices are shown in Fig.~\ref{fig:spectralSiIce}a-d. Sparse matrices are observed for amorphous silica and low-density amorphous ice, reflecting that most loops are either unlinked or connected to only a few neighbors. By contrast, $\alpha$-quartz and high-density amorphous ice exhibit much denser linking matrices, indicating that topological constraints are distributed throughout the network.

As shown in Fig.~\ref{fig:spectralSiIce}e-f, amorphous silica and low-density amorphous ice display monotonically decaying spectra characteristic of sparse and heterogeneous loop-linking networks. In contrast, $\alpha$-quartz and high-density amorphous ice develop pronounced maxima at intermediate eigenvalues, indicating a characteristic connectivity scale associated with more uniformly distributed topological constraints, with HDAI exhibiting an intermediate character between the crystalline and low-density amorphous networks. The eigenvectors of the Laplacian further reveal differences in the spatial organization of these modes through their inverse participation ratios (Fig.~\ref{fig:spectralSiIce}g). The mean IPR reveals substantially stronger mode localization in amorphous silica than in the crystalline phases or amorphous ice, indicating that the organization of topological constraints is more spatially heterogeneous in amorphous silica.

\bibliographystyle{apsrev4-2}
\bibliography{bibliography}